\documentclass[journal]{IEEEtran}
\usepackage{cite}
\ifCLASSINFOpdf
   \usepackage[pdftex]{graphicx}
\else
\fi
\usepackage{amsmath}
\usepackage{amssymb}

\usepackage{array}
\newcolumntype{M}[1]{>{\centering\arraybackslash}m{#1}}

\ifCLASSOPTIONcompsoc
  \usepackage[caption=false,font=normalsize,labelfont=sf,textfont=sf]{subfig}
\else
  \usepackage[caption=false,font=footnotesize]{subfig}
\fi

\usepackage{stfloats}
\usepackage{url}
\usepackage{soul}
\usepackage{xcolor}
\usepackage{bm}

\usepackage{multirow}

\DeclareMathOperator*{\argmin}{arg\,min}

\begin{document}
\title{Identification of TV Channel Watching from Smart Meter Data Using Energy Disaggregation}

\author{Pascal A. Schirmer, Iosif Mporas and Akbar Sheikh-Akbari% <-this % stops a space
\thanks{P. A. Schirmer and I. Mporas are with the school of Engineering and Computer Science, University of Hertfordshire, AL10 9AB Hatfield, UK (e-mail: \{p.schirmer/i.mporas\}@herts.ac.uk).}% <-this % stops a space
\thanks{Akbar Sheikh-Akbari is with the School of Built Environment, En-
gineering and Computing, Leeds Beckett University (e-mail: a.sheikh-
akbari@leedsbeckett.ac.uk).}}% <-this % stops a space}


\markboth{}%
{Shell \MakeLowercase{\textit{et al.}}: Bare Demo of IEEEtran.cls for IEEE Journals}



%\author{\IEEEauthorblockN{Pascal A. Schirmer and Iosif Mporas}
%\IEEEauthorblockA{\textit{Communications Intelligent Systems Group} \\
%\textit{School of Engineering and Computer Science}\\
%University of Hertfordshire\\
%Hatfield AL10 9AB, UK\\
%\{p.schirmer,i.mporas\}@herts.ac.uk}\\ 
%\and
%\IEEEauthorblockN{Akbar Sheikh-Akbari}
%\IEEEauthorblockA{\textit{Engineering and Computing} \\
%\textit{School of Built Environment}\\
%Leeds Beckett University\\
%Leeds LS1 3HE, UK\\
%a.sheikh-akbari@leedsbeckett.ac.uk}
%}

\maketitle

\begin{abstract}
Smart meters are used to measure the energy consumption of households. Specifically, within the energy consumption task smart meter have been used for load forecasting, reduction of consumer bills as well as reduction of grid distortions.  Except energy consumption smart meters can be used to disaggregate energy consumption on device level. In this paper we investigate the potential of identifying the multimedia content played by a TV or monitor device using the central house's smart meter measuring the aggregated energy consumption from all working appliances of the household. The proposed architecture is based on elastic matching of aggregated energy signal frames with 20 reference TV channel signals. Different elastic matching algorithms were used with the best achieved video content identification accuracy being 93.6\% using the MVM algorithm.
\end{abstract}


% \fi
\IEEEpeerreviewmaketitle

\section{Introduction}
\IEEEPARstart{O}{ver} the last decades there has been  an extensive use of smart meters in residential buildings, with 60\% of the houses in USA \cite{Cooper.2016} and 50\% of the houses in Europe \cite{ShanZhou.2017} having smart meters installed. Smart meters provide residents/consumers with information about their daily energy consumption and based on this information residents can manage or reschedule the usage of their devices to reduce electricity bills, e.g. by using some appliances like washing machines at night time during which electricity costs are usually lower \cite{Althaher.2015}.\par 

Apart from measuring household's energy consumption smart meters can also be used to provide more detailed information, as in the case of energy disaggregation where from one smart meter installed at the main inlet of the household, the usage and energy consumption on device level is extracted using Non-Intrusive Load Monitoring (NILM) methods \cite{Hart.1992}. In NILM, the aggregated signal is split into device signals using source separation methods \cite{Figueiredo.2014,Rahimpour.2017,Makonin.} or is processed by machine learning based device models to detect the existence of devices within time sliding frames \cite{Schirmer.2020b,Johnson.2013,He.2019,Harell.2019,Schirmer.2019e,Schirmer.2020c}. Specifically, variants of HMMs \cite{Makonin.2016}, CNNs \cite{Harell.2019} and LSTM \cite{Kaselimi.2019} architectures have been utilized in order to achieve accurate disaggregation. Furthermore, also elastic matching algorithms have been proven to work successfully \cite{Schirmer.2020,Liao.2014b}. By breaking down the energy consumption information on device level, the consumers can be informed about the distribution of energy consumption across home appliances and manage them, or rearrange the schedule of their operation in a more efficient way \cite{Lin.2015,Kelly.}. 

Furthermore, smart meters have been utilized for other energy related tasks, e.g. load forecasting, for reduction of consumer bills \cite{Ju.2018} or reduction of grid distortions \cite{MatthiasPilz.}. Moreover, additional information, e.g. weather condition \cite{Shimizu.2017} or socio-economic information \cite{Schirmer.2020inpress}, has been used and combined with the measurements of the smart meters. Based on NILM algorithms, smart meters can be used as non-intrusive sensors, unlike cameras and microphones, that further to energy consumption can also monitor consumers’ behaviour, device usage preferences and daily routine habits \cite{Zeifman.2012}. Therefore, in more ‘exotic’ scenarios, smart meters can be used to detect or even to predict abnormal behaviour of residents, especially in the case of elders or mental disorders \cite{Bousbiat.2020}.\par
 
However, the usage of smart meters for various tasks in consumers’ households raises the question of home data security and privacy \cite{Mrabet.2018,Anzalchi.2015}. Specifically, smart meters providing high frequency energy consumption data have raised security issues even before their major implementation in consumer households \cite{McLaughlin.2011,UrRehman.2015}. Studies have shown that even non-intrusive smart-meters enable accurate tracking of a person's location within the house, e.g. by detecting changes of lighting or other frequently used devices, or enable estimation of working routines and number of people living in a household \cite{Zhao.2018b,Dong.2017}. As these information are very personal and could even be related to security, e.g. working routines could be observed and used by criminals to plan burglaries, studies on encryption of energy data have been presented in \cite{Li.2016,Papagiannakopoulou.2013,Wang.2011}. \par 

\begin{figure*}[ht]
\centering
\includegraphics[width=\textwidth]{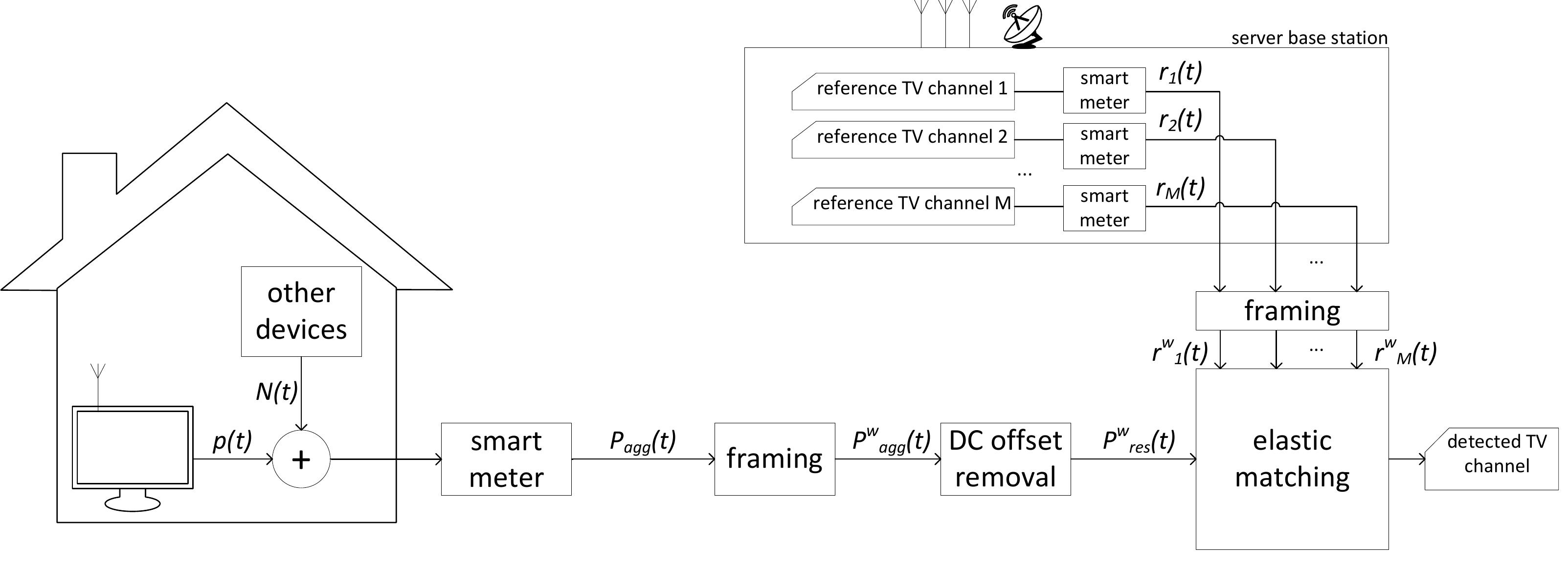}
\caption{Block diagram of the architecture for TV channel watching identification using smart meter data.}
\label{fig: architecture}
\end{figure*}

Multimedia content identification from energy consumption has been presented in \cite{UlrichGreveler.2012} where the study was limited to measurements of correlations of energy predictors in isolated monitors (intrusive load monitoring) on specific video chunks. The application of NILM techniques, for identifying the use of appliances during a time sliding frame in a household, has been reported in literature utilizing high \cite{Bouhouras.2019,Gao.2015} and low frequency features \cite{Schirmer.542020582020,Schirmer.542020582020b}. However, to the best of the authors’ knowledge, the application of NILM in recognising which TV channel is being viewed, by just observing the aggregated energy consumption signal, has not been investigated. In this paper, we investigate the potential of using smart meter's measurement data to identify the viewed TV channels. Specifically, given that the smart meter is measuring the aggregated energy consumption from all operating appliances of a household, we investigate the possibility of identifying the TV channel a resident is watching from the aggregated energy consumption signal.\par 

The remainder of this paper is organized as follows: In Section \ref{sec: TV Channel Watching Identification from Smart Meter Data Architecture} the proposed architecture for detection of TV channel watching from the central household's smart meter is presented. In Sections \ref{sec: Experimental Setup} and \ref{sec: Experimental Results} the experimental setup and the evaluation results are presented, respectively. Finally, the paper is concluded in Section \ref{sec: Conclusion}.

 

%\subsection{Subsection Heading Here}
%Subsection text here.

%\IEEEpubidadjcol

%\subsubsection{Subsubsection Heading Here}
%Subsubsection text here.

\section{TV Channel Watching Identification from Smart Meter Data Architecture}\label{sec: TV Channel Watching Identification from Smart Meter Data Architecture}
The presented architecture deems to investigate the potential of identifying the TV channel watching preferences of residents using the aggregated energy consumption signal acquired outside the house from a smart meter installed at the main inlet of the household. The conceptual diagram of the architecture for identification of watched TV channels using explicitly smart meter’s energy data is illustrated in Fig. \ref{fig: architecture}. Specifically, the architecture in Fig. \ref{fig: architecture} underlies the following five assumptions:

\begin{enumerate}
\item The number of TV channels is of medium size ($\sim$ 20 different channels).

\item The noise of the ‘other devices’ is simulated through multiple different scenarios, with different noise levels each.

\item There is no time lag between the recordings in the household and the server base station.

\item In the considered household a maximum of one TV device is turned on at the same time.

\item The TV operates in real-time watching mode not in video on demand mode. 
\end{enumerate}

As can be seen in Fig. \ref{fig: architecture} a smart meter is measuring the aggregated energy consumption, $P_{agg}(t)$, of a household. The aggregated signal is the sum of the energy consumption of all the devices of the house and in the present setup we consider the TV device (or a monitor) as the target device with energy consumption $p(t)$ and all other home appliances having energy consumption $N(t)$, i.e.
\begin{equation}\label{Eq. Baseline}
P_{agg}(t) = p(t) + N(t) = p(t) + \sum_{i=1}^{N-1}n_i(t)
\end{equation}
where $N$ is the number of all appliances of the household, including the TV device, e.g. fridge, washing machine, etc., operating in the considered household.

Subsequently, the aggregated signal, $P_{agg}(t)$, is frame blocked in frames of constant length equal to $W$ samples and from every energy frame, $P_{agg}^w(t) \in \mathbb{R}^w: P_{agg}(t-W:t)$, the $DC$ offset is removed, resulting to $P_{res}^w$. The reason for the $DC$ offset removal is the fact that most of the common home appliances like fridges, refrigerators, boilers, electric heating bodies, electric ovens etc., consume energy at the order of 200-2000 Watts while the average energy consumption of a TV device or monitor is at the order of 25-250 Watts. Therefore, the main part (DC part) of the energy consumption signal within each frame will come from devices with high energy consumption and by removing it in the remaining residual signal, $P_{res}^w (t)\in \mathbb{R}^W$, the contour shape characteristics of the energy signal of devices with lower energy consumption like the TV device will be shown more clearly.

Except the household aggregated energy consumption measurements, we consider a server base station where the broadband signals from $M$ TV channels are received, assuming that $M$ are all available TV channels. Each of the received signals is played by $M$ reference TV devices of the same brand and model and the corresponding energy consumption signals, $r_m (t)$, with $1 \leq m \leq M$, are measured by smart meters. It is worth mentioning that the TV devices used at the server base station are not the same with the TV device of the household. Next, each of the $M$ reference signals, $r_m (t)$, is frame blocked in frames of constant length equal to $W$ samples, i.e. $r_m^w (t) \in \mathbb{R}^W$.

\begin{figure*}[ht]
\centering
\includegraphics[width=\textwidth]{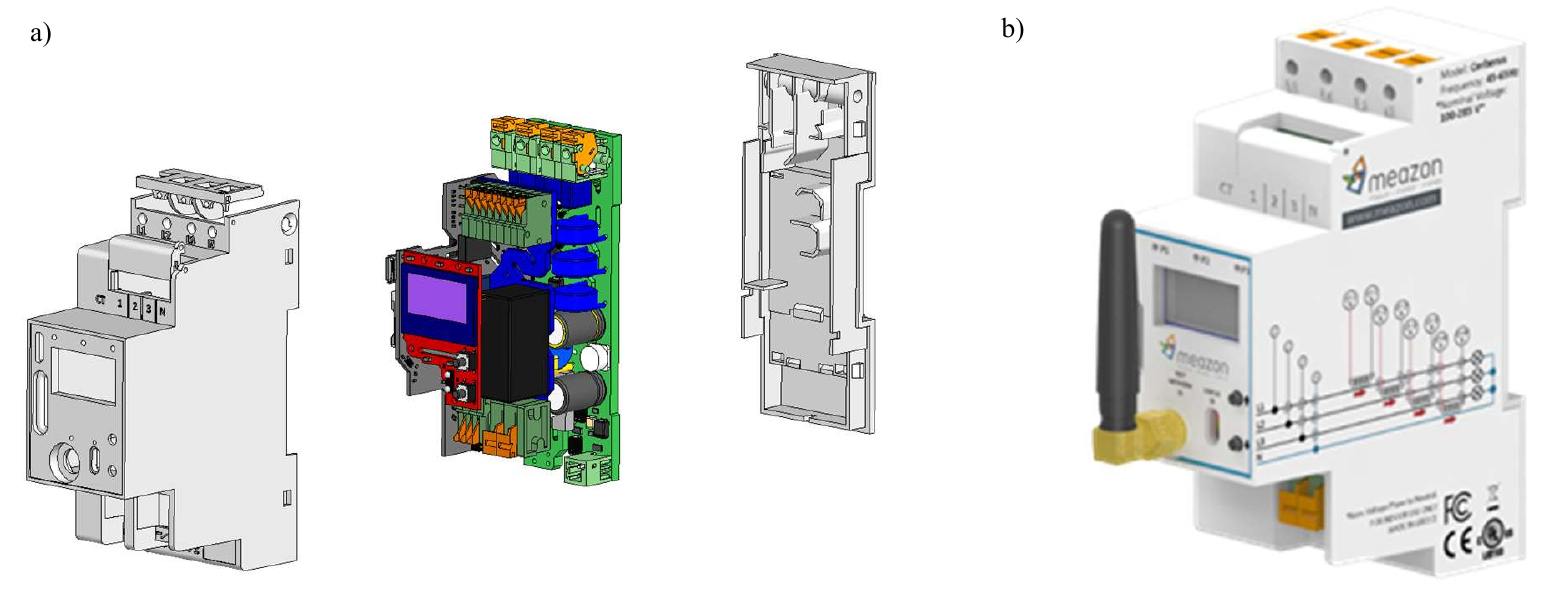}
\caption{Cerberus smart meter from MEAZON: a) CAD drawing including PCB and casing b) smart meter device.}
\label{fig: smartMeter}
\end{figure*}

Assuming no significant time-lag at the reception of the TV broadband signal between the house and the server base station template matching (elastic matching) is performed between the unknown residual signal, $P_{res}^w (t)$, and each of the $M$ reference signals, $r_m (t)$, and the reference TV channel energy consumption signal with the lowest distance from the residual household’s signal is considered as the detected TV channel $Ch$ watched at time frame $t$, i.e. 
\begin{equation}\label{Eq. Solution 1}
Ch(t) = \argmin_{1 \leq m \leq M} \{f(P_{res}^w(t),r_m(t))\}
\end{equation}
where $f()$ is a template (elastic) matching algorithm and $Ch(t) \in [1,...,M]$ is the detected TV channel watched at time frame $t$.

In order to provide more distinctive information, the above described architecture can be expanded with the elastic matching algorithm $f()$ not only considering the aggregated active power samples, but a set of $F$ features $X_{res}^w (t) \in \mathbb{R}^{FxW}$ and $r_m^{'}(t) \in \mathbb{R}^{FxW}$. The $F$ features are computed from the raw current and voltage samples, $i_{agg}(t)$ and $v_{agg}(t)$, providing set of most common electrical features, e.g. reactive power or harmonics, and include the active power samples. Thus Eq. \ref{Eq. Solution 2} can be rewritten as:
\begin{equation}\label{Eq. Solution 2}
Ch(t) = \argmin_{1 \leq m \leq M} \{f(X_{res}^w(t),r_m^{'}(t))\}
\end{equation}

In the presented architecture the case of not watching TV can be considered either by applying threshold values to the elastic matching algorithms of the detected TV channels or by considering not watching TV as an additional class, i.e. the ($M$+1)$^{th}$ one. Also, the architecture can be expanded to the case of more than two TV devices to play different channels simultaneously by considering one of them as belonging to the ‘other devices’ or by selecting all elastic matching scores below a threshold as detected devices.

\section{Experimental Setup}\label{sec: Experimental Setup}
The video classification architecture presented in Section \ref{sec: TV Channel Watching Identification from Smart Meter Data Architecture} was evaluated using the dataset, energy features, classification models and experimental protocols presented below.

\subsection{Evaluation Data}\label{sec: Evaluation Data}
As there is no benchmark dataset with energy consumption measurements from TV or monitor devices we recorded a dataset for video classification using the publicly available video footage from the THUMOS challenge 2014 \cite{THUMOS14}. In detail, we selected the first 26 videos from the background data, excluding videos 11, 14, 20, 23, 24 and 25 as their duration was less than 1 min, resulting in a set of 20 videos. Energy consumption data measurements were taken from two different monitors, namely an Acer P235H which was used as the test screen at the house as shown in Fig. \ref{fig: architecture} and an Iiyama Prolite B2483HS which was used as reference pattern at the servers station. The technical characteristics of the two monitors are tabulated in Table \ref{Table: screens}.  

\begin{table}[ht]
\small
\renewcommand{\arraystretch}{1.0}
\caption{Technical characteristics of the monitors used}
\label{Table: screens}
\centering
\begin{tabular}{ccc}
\hline
& \textbf{Acer P235H}& \textbf{Iiyama B2483HS} \\
\hline
Technology      		& LCD & LED \\
Screen size (inch)		& 23  & 24  \\
Brightness (cd/$m^2$)  & 300 & 250 \\
Resolution (pixels)    & 1920x1080 & 1920x1080 \\
Power (Watts)    & 31.7 & 24.9 \\
\hline
\end{tabular}
\end{table}

For measuring the electrical energy consumption of the monitors, a hardware prototype smart meter developed by MEAZON (https://meazon.com/) was used for the first 1 minute of each video. The smart meter is a energy circuit-level meter, measuring current, voltage, line frequency, active and reactive power and energy as well as harmonics and crest factor. It was designed for monitoring loads in an electrical board in commercial or industrial buildings or homes with an internal control capability up to 16 Amperes that can be extended further by driving an external relay. The USB port and the ARM Cortex M3 CPU run at 32 MHz Clock Speed with 512 Kbyte of In-System Programmable Flash and 32 Kbytes of Memory. All the above parameters, e.g. active power or reactive power, were outputted once per electrical cycle by the smart meter, thus at a rate of 50 times per second. Specifically, the internal sampling rate for calculating each feature is 8000 Hz, thus for each output sample 160 samples have been used for the calculation.  The prototype of the used smart meter is shown in Fig. \ref{fig: smartMeter}.

As regards the UK-DALE database \cite{Kelly.2015b}, it was used to generate the energy consumption signal of $N$-1 ‘other devices’ from a real house. In particular, UK-DALE was chosen among other online available databases, e.g. REDD \cite{Kolter.2011}, ECO \cite{Beckel.2014} or AMPds \cite{S.MakoninF.PopowichL.BartramB.Gill.2013}, since it provides high frequency measurements (16 kHz) of households combined with a large number of appliances operating in parallel. In specific, we used one hour of the first house of UK-DALE (maximum of 52 appliances) from the $2^{nd}$ of December 2014 between 4pm-5pm due to the presence of 1 up to 26 electrical appliances within this time window, with none of them being a TV device. In order to have the same sampling rate as measured by the smart meter the data were down-sampled to 50 Hz.

\subsection{Feature Extraction and Feature Ranking}\label{sec: Feature Extraction and Feature Ranking}
\begin{figure}[ht]
\centering
\includegraphics[width=3.5in]{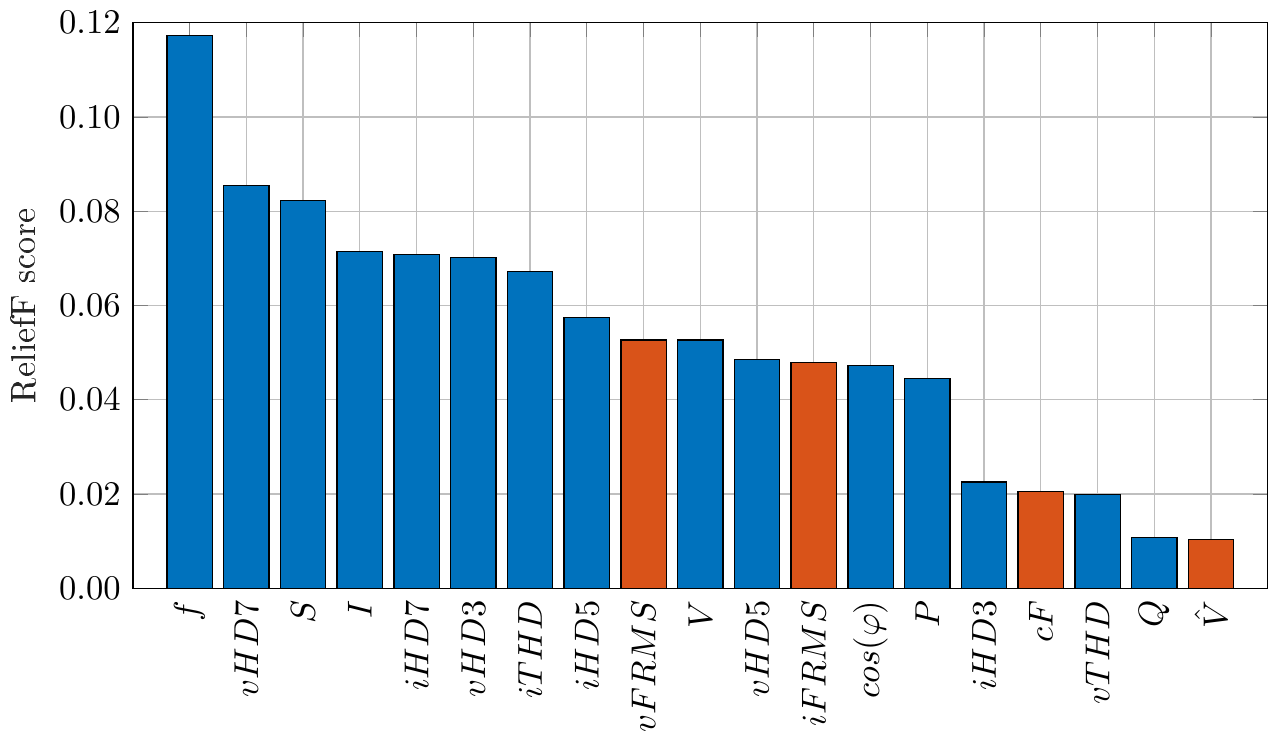}
\caption{Feature ranking for the set of 4 statistical (red) and 15 electrical features (blue).}
\label{fig: featureRanking}
\end{figure}

The raw samples of the aggregated current and voltage $i_{agg}$ and $v_{agg}$ was frame blocked in frames of 160 samples. For every frame a feature vector was calculated consisting of 4 statistical features (peak voltage ($\hat{V}$), rms current/voltage (iFRMS, vFRMS), crest factor of current (cF)) and 15 electrical features (current (I), voltage (V), active power (P), frequency (f), reactive power (Q), apparent power (S), load angle ($\varphi$), total harmonic distortion current/voltage (iTHD, vTHD), $3^{rd}$/$5^{th}$/$7^{th}$ harmonic current/voltage (iHD3/5/7, vHD3/5/7) resulting to feature vectors of dimensionality equal to $F=19$. In order to calculate the statistical importance of the 19 features the ReliefF feature ranking algorithm \cite{Urbanowicz.} was used by averaging the ranking scores across the 20 measured video signals. The results are illustrated in Fig. \ref{fig: featureRanking}.

\begin{figure}[ht]
\centering
\includegraphics[width=3.5in]{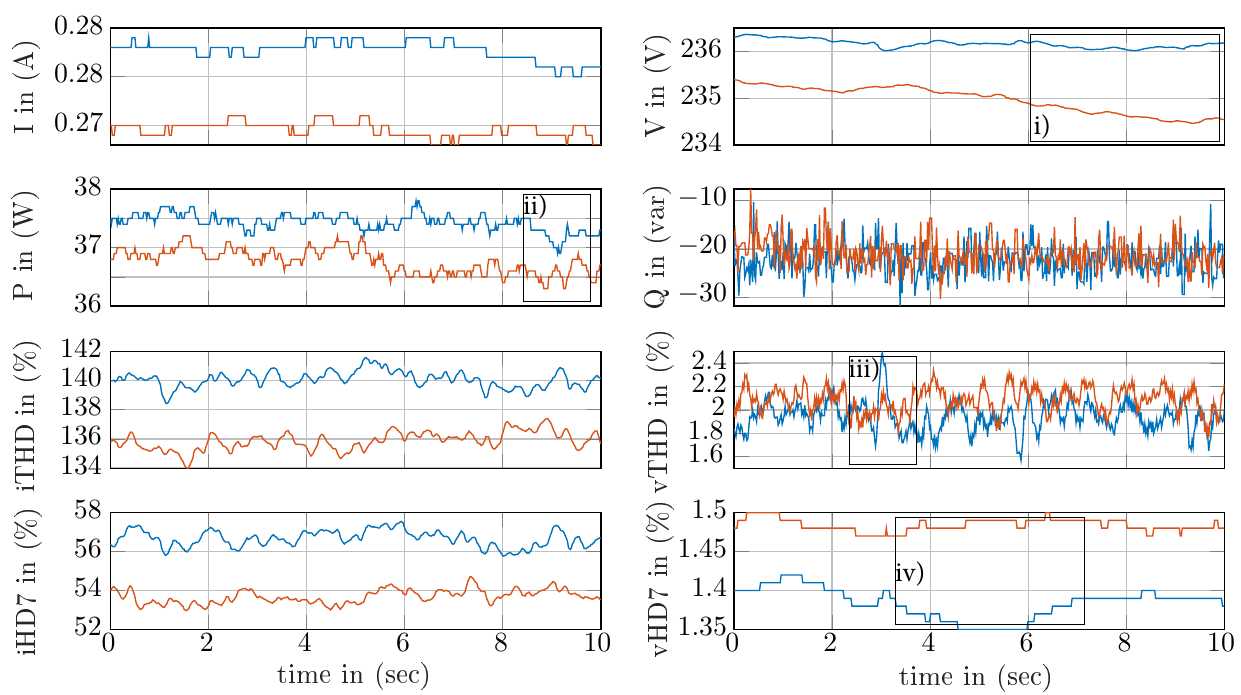}
\caption{Illustration of eight different features for two different video sequences on the same monitor (blue: video \#1, red: video \#2)}
\label{fig: features two singals}
\end{figure}

As can be seen in Fig. \ref{fig: featureRanking} most of the electrical features in general outperform the statistical features with the most dominant features being the frequency, apparent power, raw current as well as of set current and voltage harmonics. This is in line with previous publications reporting high importance of electrical features \cite{Ghorbanpour.2018,Huang.2020}. Regarding current and voltage harmonics, they carry significant amount of the video playing energy signal's information, e.g. both $iHD5$ and $iHD7$ have high feature ranking scores. Active power $P$, reactive power $Q$ and apparent power $S$ carry similar information as they can be computed by the relation $S=\sqrt{P^2 + Q^2}$, thus $Q$ has a relatively low feature ranking score as it can be computed by $S$ and $P$. Furthermore, as it is not clear how discriminative electric measurements of monitors are when being used for TV channel classification, e.g. electrical measurements might be filtered through a large capacitor at the monitor input, the time domain signals of eight different features are illustrated for two different video sequences in Fig. \ref{fig: features two singals}.

As illustrated in Fig. \ref{fig: features two singals} all eight features show different shapes for the two different video signals respectively. Specifically, the following four instances marked with bounding boxes i) - iv) are analysed. In specific, the first case denoted as '(i)' illustrates the voltage envelop over time for both videos showing a significant stronger decrease of the red curve compared to the blue curve. The second case denoted as '(ii)' illustrates the difference between the two signals for the active power consumption. In detail, the time envelop of the signals is inverse showing a decrease in the blue signal and a increase in the red signal. The third case denoted as '(iii)' illustrates the envelop of the THD of the voltage showing a significant peak for the blue signal while the red signal is relatively constant. The last case '(iv)' shows the time envelop of the $7^{th}$ voltage harmonic, with the red signal being relatively constant while the blue signal has a significant drought. To summarize, Fig. \ref{fig: features two singals} illustrates that especially features with high feature ranking showing significantly different patterns in the time domain for two different videos.

\begin{figure}[ht]
\centering
\includegraphics[width=3.5in]{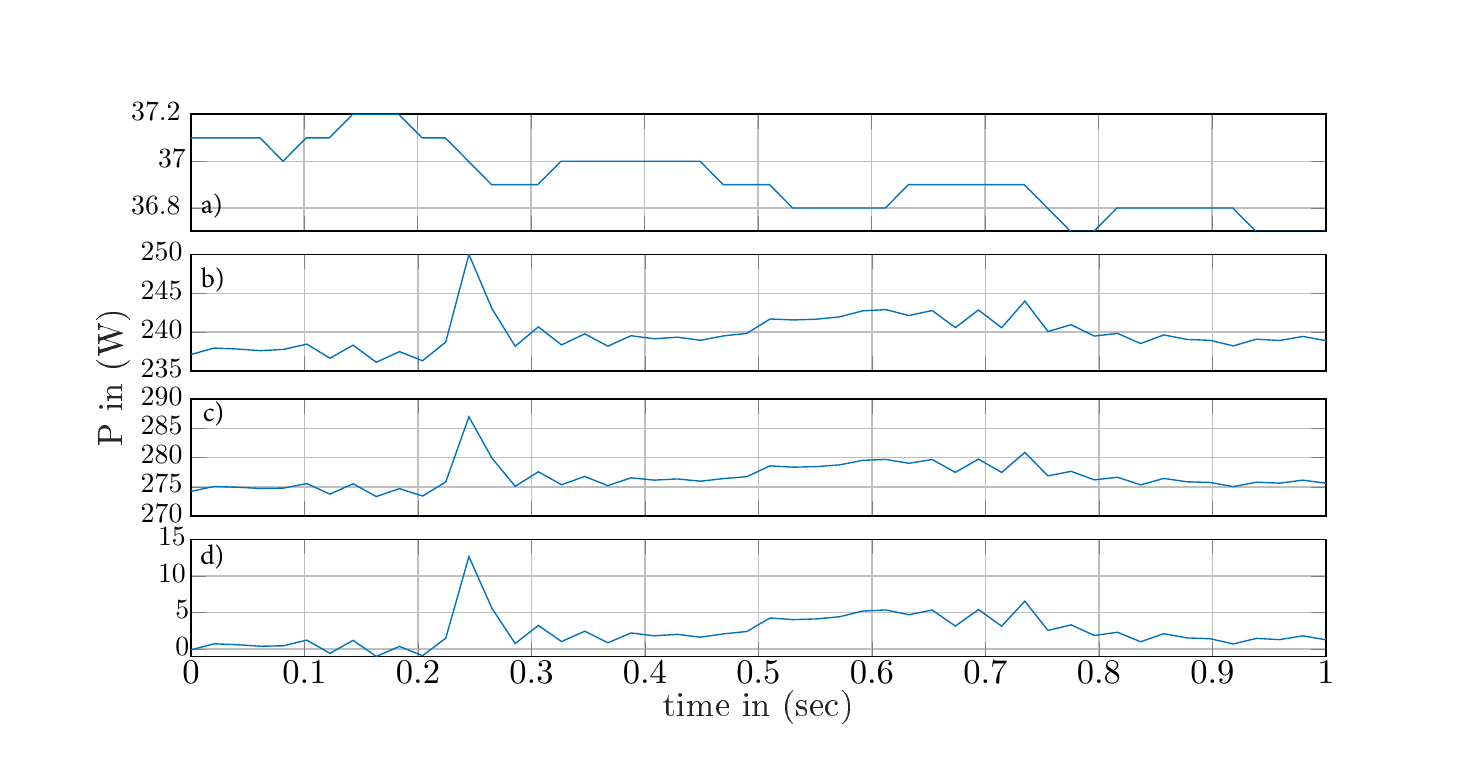}
\caption{Illustration of the active power of (a) the TV/monitor only, (b) other appliances, (c) the aggregated signal and (d) the aggregated signal after DC offset removal.}
\label{fig: time domain signal P}
\end{figure}

\begin{figure}[ht]
\centering
\includegraphics[width=3.5in]{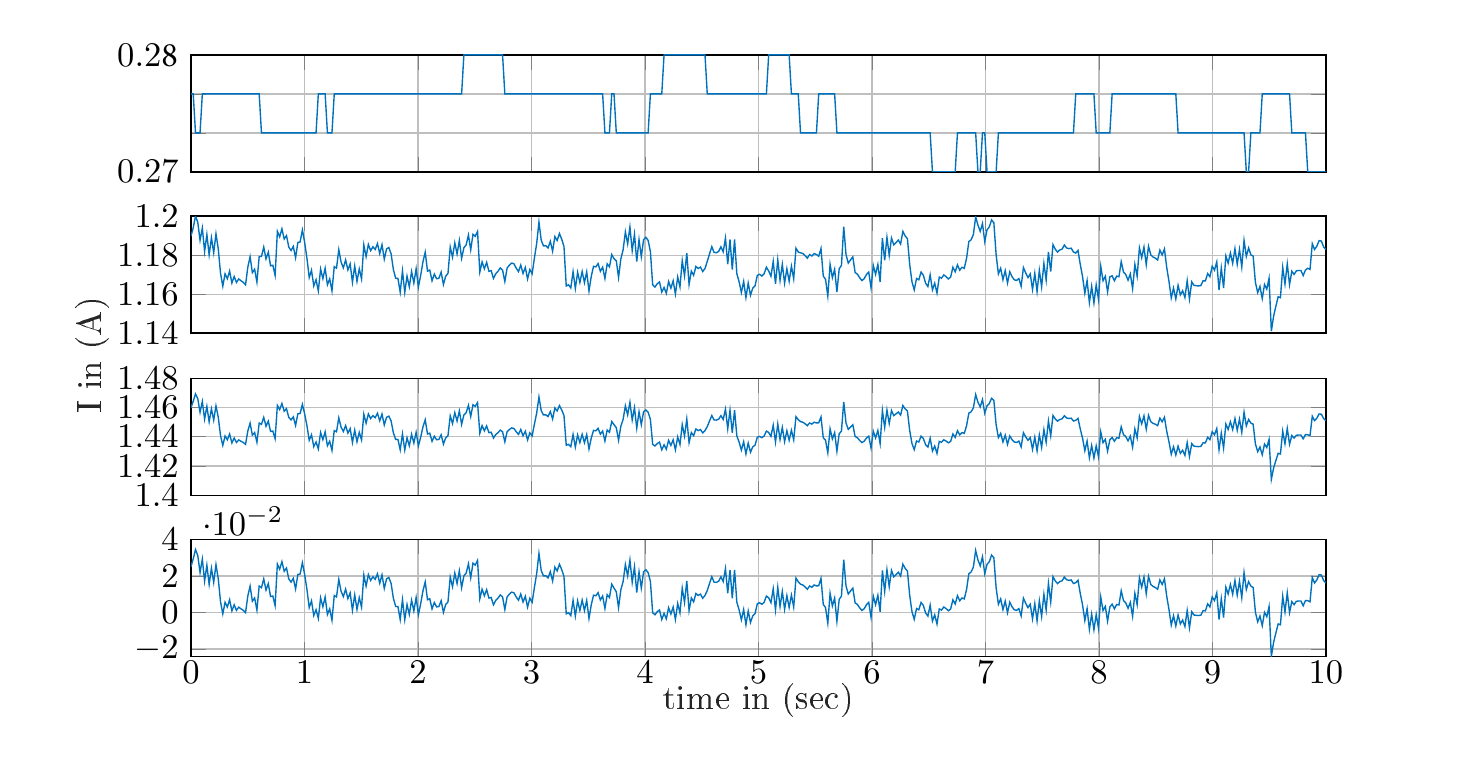}
\caption{Illustration of the current of (a) the TV/monitor only, (b) other appliances, (c) the aggregated signal and (d) the aggregated signal after DC offset removal.}
\label{fig: time domain signal I}
\end{figure}

\begin{figure}[ht]
\centering
\includegraphics[width=3.5in]{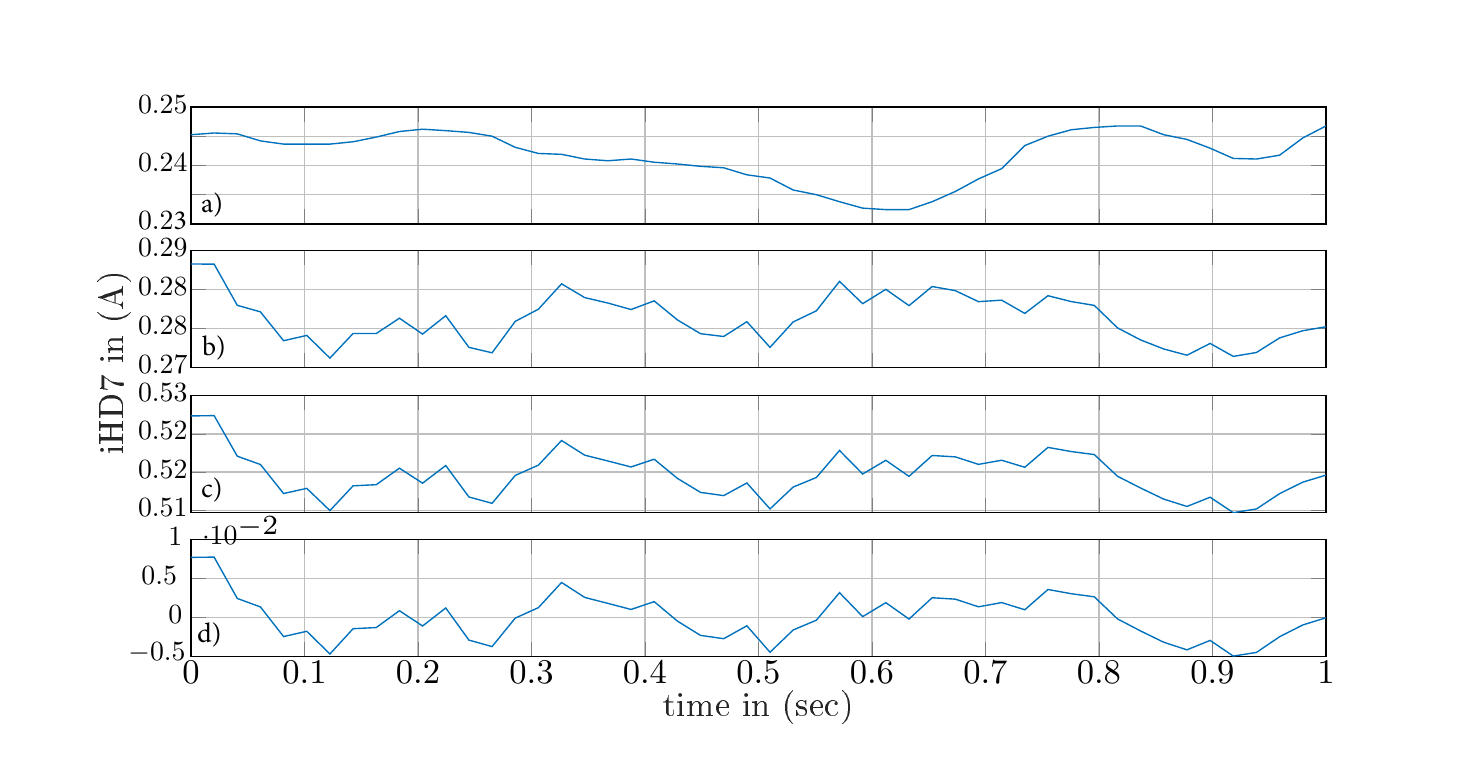}
\caption{Illustration of the $7^{th}$ current harmonic of (a) the TV/monitor only, (b) other appliances, (c) the aggregated signal and (d) the aggregated signal after DC offset removal.}
\label{fig: time domain signal I7}
\end{figure}

Further to feature ranking measurements, examples of the monitor’s energy consumption information carried by the active power, raw current and the $7^{th}$ current harmonic are illustrated in Fig. \ref{fig: time domain signal P} - Fig. \ref{fig: time domain signal I7}, respectively, including the aggregated signals before and after frame's DC offset removal.

\subsection{Elastic Matching Algorithms}\label{sec: Elastic Matching Algorithms}
For the classification, four different well known and widely used elastic matching algorithms were employed, namely the Dynamic Time Warping (DTW), soft Dynamic Time Warping (sDTW), Multi Variance Matching (MVM), and Global Alignment Kernel (GAK) \cite{Cuturi.2017,Cuturi.2011,LonginJanLatecki.2007}. The mathematical descriptions of these four template matching algorithms are presented below.

Lets considering the aggregated power consumption signal $P_{agg}(t) \forall t:t \in \{1,\cdots,T\}$ as measured by a smart meter. For simplicity let the $w^{th}$ frame of $P_{res}^w$ be denoted by $P_a$, where $P_a=[p(i),p(i+1), \cdots, p(i+K)]$ be a sequence of length $K$ where $p(i)$ is the $i^{th}$ sample of $P_{agg}$. Furthermore, let $P_b=[p(j),p(j+1), \cdots, p(j+L)]$ be a second sequence of length $L$ where $p(j)$ is the $j^{th}$ sample of $P_{agg}$ and $K<L$. Furthermore, let $\Delta(P_a,P_b)=[\delta(p_a^k,p_b^l)]_{i,j} \in \mathbb{R}^{KxL}$ be an arbitrary cost matrix, where $\delta(\cdot)$ is a distance metric e.g., Euclidean distance, Manhattan distance or Kullback-Leibler (KL) distance and $\langle A,\Delta(P_a,P_b)\rangle$ being the inner product of matrix $A$ with the cost matrix $\Delta(P_a,P_b)$, where $A$ is an alignment matrix with $A_{k,l}$ giving the scores of $A$. 

Based on the above the generalized $min^{\gamma}$ operator, with the smoothing parameter $\gamma \ge 0$ can be written as in Eq. \ref{Eq: sdtw} and is referred to as soft dynamic time warping $dtw_{\gamma}$.

\begin{equation}\label{Eq: sdtw}
dtw_{\gamma}:= min^{\gamma}\{\langle A,\Delta(P_a,P_b)\rangle\, A \in A_{k,l}\}
\end{equation}
\begin{equation}\label{Eq: sdtw update}
min^{\gamma}\{a_1, \cdots, a_k\}:= \begin{cases}
      min_{i \le k} a_i & \gamma=0\\
      -\gamma log \sum_{i=1}^{k} e^{-a_i/\gamma} & \gamma > 0
    \end{cases}
\end{equation}
where the original DTW score is recovered by setting $\gamma=0$.

Extending the previous definition of sDTW the Global Alignment (GA) kernel is defined as the exponentiated soft-minimum of all alignments distances and can be written as in Eq. \ref{Eq: gak}

\begin{equation}\label{Eq: gak}
k_{GA}^{\gamma}:= \sum_{A \in A_{k,l}} e^{-\langle A,\Delta(P_a,P_b)\rangle/\gamma}
\end{equation}
where $\gamma>0$ is the smoothing parameter of the kernel. Compared to DTW, $k_{GA}^{\gamma}$ incorporates the whole spectrum of costs $\langle A,\Delta(P_a,P_b)\rangle$ and thus provides a richer representation than the absolute minimum of set $A$, as considered by DTW.

In contrast to DTW, sDTW and GAK, MVM tries not to find the optimal alignment between the two sequences $P_a$ and $P_b$, but also considers the alignment of subsequences. Therefore, MVM tries to find a subsequence $P_a^{'}$ of length $K$ such that $P_b$ best matches  $P_a^{'}$. To formally describe MVM the difference matrix $r$ between the two sequences $P_a$ and $P_b$ and is defined as follows:

\begin{equation}\label{Eq: distance r}
r=(r_{kl}) = (p_a^k - p_b^l)
\end{equation}
Furthermore, $r_{kl}$ is treated as a directed graph with the following links \cite{LonginJanLatecki.2007}:
\begin{equation}\label{Eq: distance graph}
r_{kl} \leftrightarrow r_{rs} \; \; \; with \; r-k=1 \; and \; l+1 \le l + K-L
\end{equation}
Using Eq. \ref{Eq: distance r} and Eq. \ref{Eq: distance graph} the least-value path in terms of the linkcost and pathcost can be written as described in \cite{LonginJanLatecki.2007}.

The free parameters of each elastic matching algorithm were empirically optimized after grid search on a bootstrap training subset utilizing 50\% of the evaluation data. The grid search results are shown in Table \ref{tab: System tuning} and the best elastic matching classification accuracy corresponding to the optimal values of each elastic matching algorithm are shown in bold. 

\begin{table}[ht]
\renewcommand{\arraystretch}{1.0}
\caption{Classification accuracy (\%) for different parameter values of $sDTW$, $GAK$ and $MVM$ algorithms.}
\centering
\begin{tabular}{ccccccc}
\hline
\multicolumn{7}{c}{\textbf{sDTW}}            \\
\hline
\centering  \bm{$\gamma$} & \textbf{1}  & \textbf{2} & \textbf{5} & \textbf{10} & \textbf{100} & \textbf{500} \\
\hline
\centering      &   91.0\% & 91.1\%          & \textbf{91.3}\%   & 90.1\%  & 89.8\% & 89.8\% \\
\hline
\multicolumn{7}{c}{\textbf{GAK}}            \\
\hline
\centering  \bm{$\gamma$} & \textbf{1}  & \textbf{2} & \textbf{5} & \textbf{10} & \textbf{100} & \textbf{500} \\
\hline
\centering      &   52.3\% & 65.9\%  & \textbf{71.8\%} & 71.4\%  & 69.7\% & 63.2\% \\
\hline
\multicolumn{7}{c}{\textbf{MVM}}            \\
\hline
\centering \textbf{v} & \textbf{5}  & \textbf{10} & \textbf{15} & \textbf{20} & \textbf{25} & \textbf{30} \\
\hline
\centering      &   95.5\% & \textbf{95.6\%} & 95.5\%  & 95.5\%  & 95.5\% & 95.5\% \\
\hline
\end{tabular}\label{tab: System tuning}
\end{table}

As can be seen in Table \ref{tab: System tuning} the optimal parameters of the elastic matching algorithms used are $\gamma=5$ for $sDTW$, $\gamma=5$ for $GAK$ and $v=10$ for $MVM$. The best classification accuracy on the bootstrap training data was achieved by $MVM$ algorithm and was equal to 95.6\% outperforming all other evaluated elastic matching algorithms.

\subsection{Experimental Protocols}\label{sec: Experimental Protocols}
To evaluate the proposed architecture three different experimental protocols were utilized:
\\
\\
(A) evaluation under a noiseless conditions, i.e. $N(t)=0$, was carried out to determine whether or not different videos can be distinguished from their electrical energy signals recorded from the same monitors, thus in this protocol it will be $P(t)=P^{'}(t)$;
\\
\\
(B): evaluation with additional ‘other devices’ was carried out, thus $N(t)=\sum_{i=1}^{N-1} n_i(t)$, using the aggregated energy signal from the UK-DALE dataset and reference patterns from the same monitor, i.e. $P(t)=P^{'}(t)$;
\\
\\
(C): evaluation with additional ‘other devices’ was carried out using different monitors, i.e. $P(t) \neq P^{'}(t)$, using Acer P235H monitor at the target house and Iiyama B2483HS monitor at server station. 
\\
\\
For the additional ‘other devices’ 59 different sets of recordings were randomly selected from the 1 hour duration measurements of the UK-DALE dataset and added to the energy measurement of the respective video signal. It must be noted that 59 noise scenarios were chosen in order to avoid zero padding for the $60^{th}$ noise scenario, as the 1 hour UK-DALE datafile is slightly shorter than 60 minutes. 

\section{Experimental Results}\label{sec: Experimental Results}
The architecture presented in Section \ref{sec: TV Channel Watching Identification from Smart Meter Data Architecture} for the identification of the TV channels watched using an outdoors smart meter was evaluated according to the experimental setup described in Section \ref{sec: Experimental Setup}. The performance of the three evaluated protocols was estimated in terms of accuracy ($ACC$) and in terms of F-score ($F_1$), i.e. 
\begin{equation}\label{Eq. ACC}
ACC = \frac{TP+TN}{TP+TN+FP+FN}
\end{equation}
\begin{equation}\label{Eq. F1}
F_1 = \frac{2 \cdot TP}{2 \cdot TP + FN + FP}
\end{equation}
where $TP$ are the true positives, $TN$ are the true negatives, $FP$ are the false positives and $FN$ are the false negatives, respectively. For each of the three experimental protocols (A, B and C) 21 energy signals (from playing 20 videos and one experiment with no video played) were tested for 60 different noise scenarios and the averaged results are tabulated in Table  in terms of $ACC$ and $F_1$ scores for the noiseless (A), noisy (B) and noisy using different monitors (C) experimental protocol.

\begin{table}[ht]
\renewcommand{\arraystretch}{1.0}
\caption{Classification results (\%) for three experimental protocols (A) noiseless, (B) noisy and (C) noisy using different monitors, averaged over 60 different noise scenarios}
\centering
\begin{tabular}{c|ccc|ccc}
\multirow{2}{*}{\textbf{Classifier}} & \multicolumn{3}{c|}{\textbf{ACC}} & \multicolumn{3}{c}{\textbf{F1}} \\ \cline{2-7}
                & \textbf{A}   & \textbf{B} & \textbf{C} & \textbf{A}  & \textbf{B}        & \textbf{C}  \\ \hline \hline
\textbf{DTW}                                  & 100.0      & 82.6     & 81.1     & 100.0     & 81.6     & 80.2     \\
\textbf{sDTW}                                 & 100.0      & 89.3     & 87.1     & 100.0     & 88.4     & 86.0     \\
\textbf{GAK}                                  & 100.0      & 67.2     & 63.7     & 100.0     & 66.4     & 62.8     \\
\textbf{MVM}                                  & 100.0      & \textbf{94.7}     & \textbf{93.8}     & 100.0     & \textbf{94.3}     & \textbf{93.3}    \\ 
\end{tabular}\label{tab: results}
\end{table}

As can be seen in Table \ref{tab: results} all four elastic matching algorithms were able to identify the played videos with 100\% accuracy when intrusive load monitoring was used (protocol A), thus under a noiseless scenario. When identification was performed using the aggregated signal (protocol B) and using the same monitor (Acer P235H), $MVM$ outperformed all other elastic matching algorithms achieving accuracy of 94.7\% and $F_1$ score 94.3\%. In protocol C, elastic matching was performed on the signals from different monitors and $MVM$ again achieved the highest performance among all evaluated algorithms (accuracy 93.8\% and  $F_1$ score 93.3\%), which is in agreement with our previous study \cite{Schirmer.2020} where $MVM$ was also found to perform well on the $NILM$ task.

\begin{figure*}[!t]
\centering\subfloat[Protocol (A) noiseless (DTW)]{\includegraphics[width=0.33\textwidth]{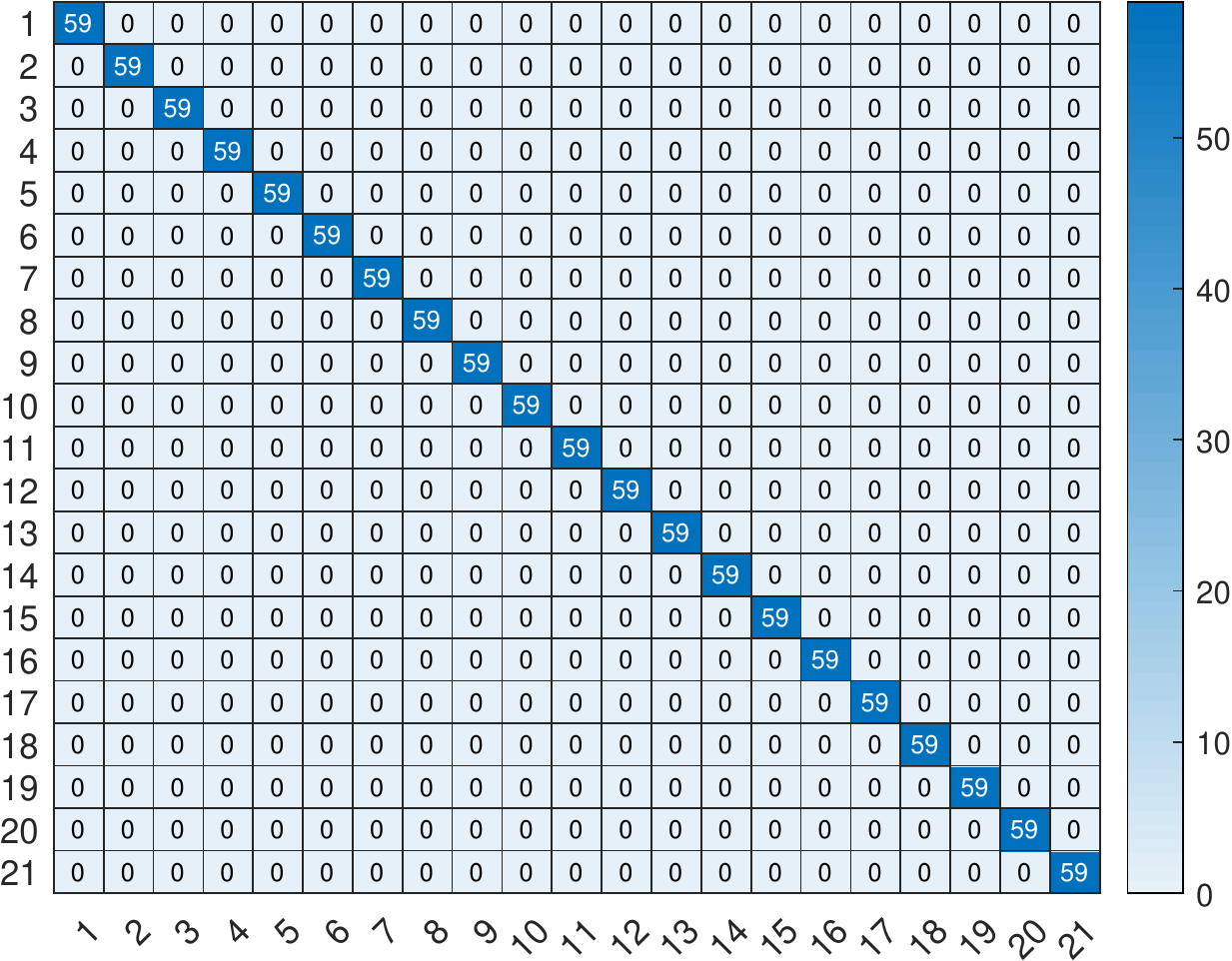}
\label{fig: DTW A}}
\subfloat[Protocol (B) noisy (DTW)]{\includegraphics[width=0.33\textwidth]{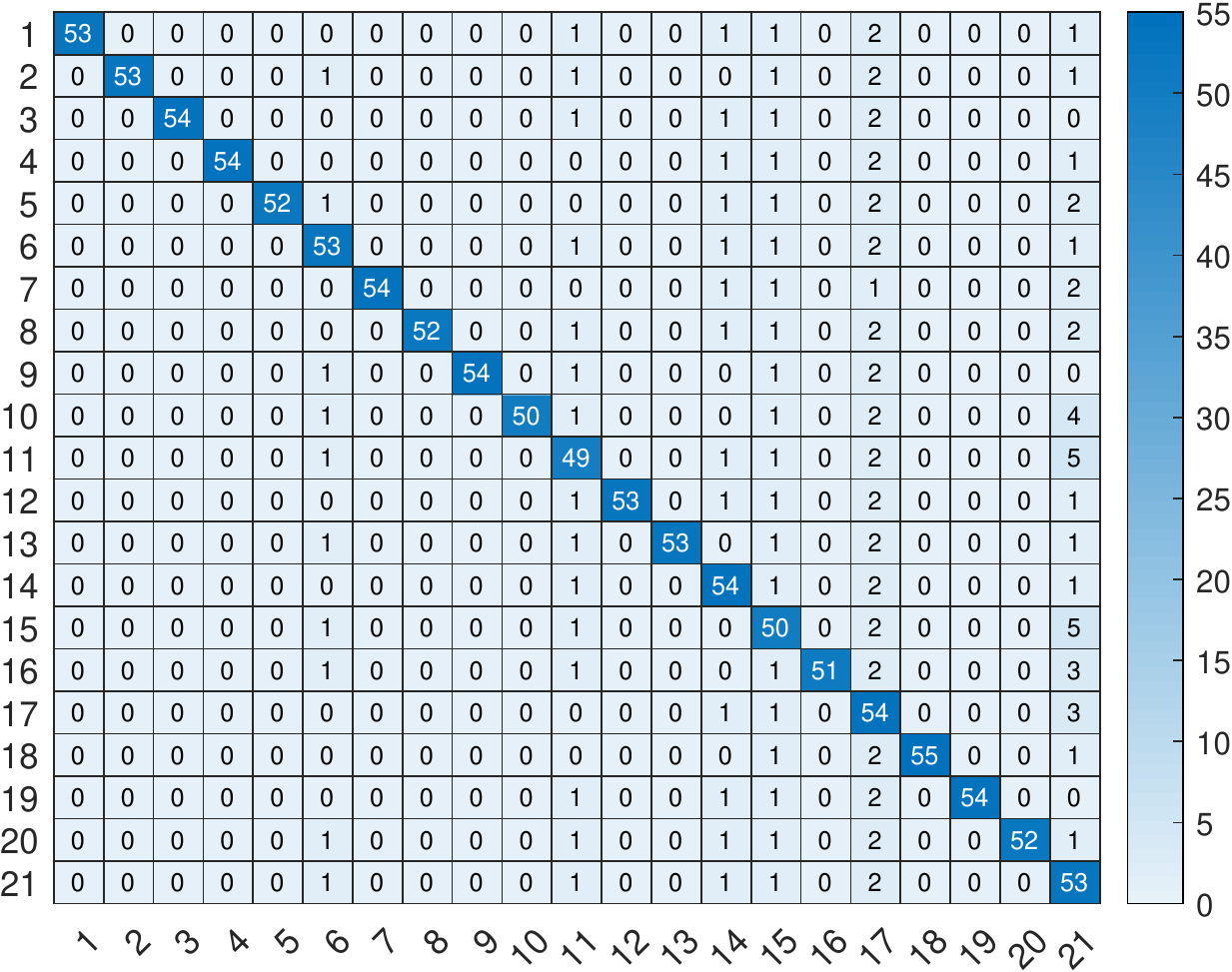}
\label{fig: DTW B}}
\subfloat[Protocol (C) noisy different monitor (DTW)]{\includegraphics[width=0.33\textwidth]{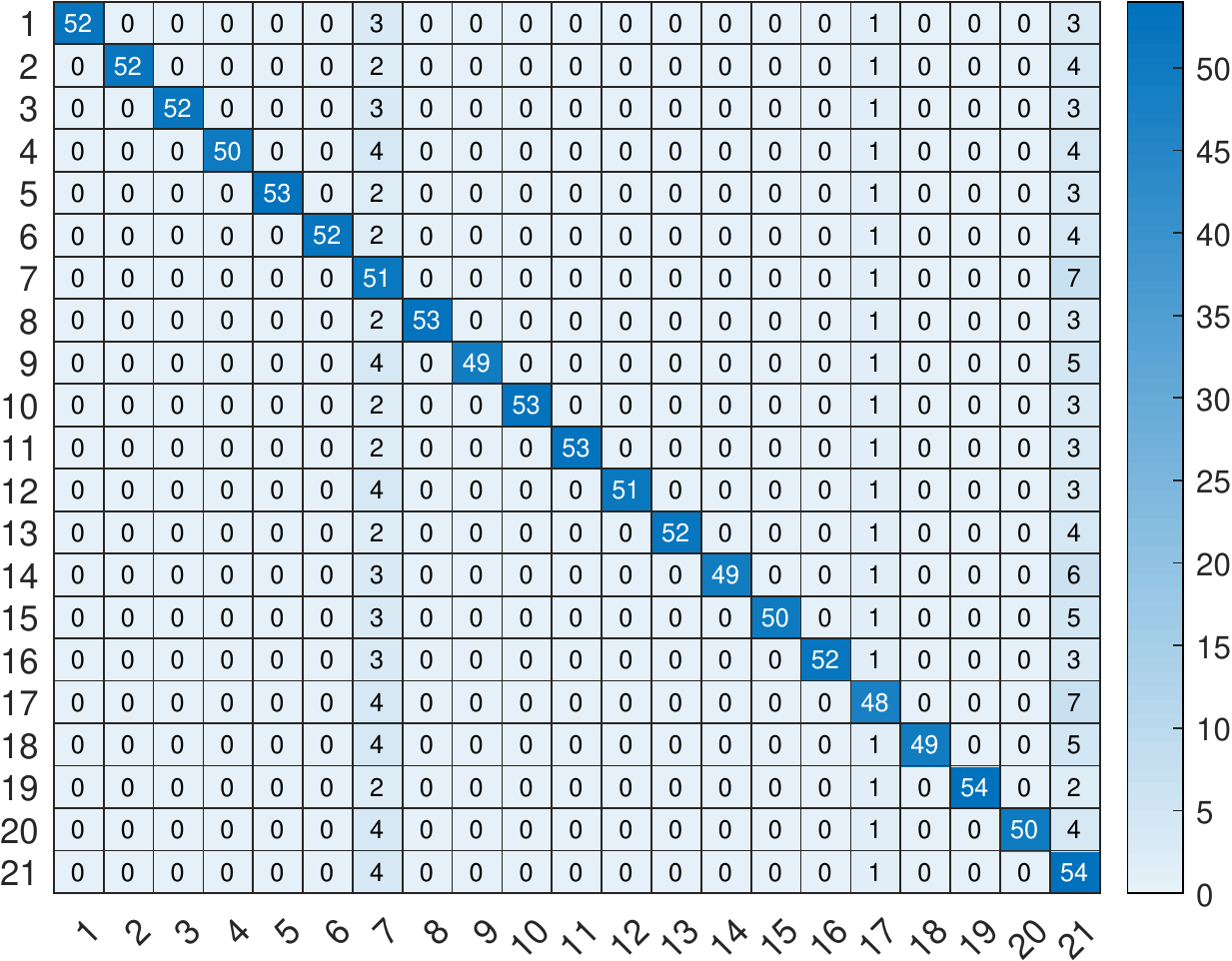}
\label{fig: DTW C}} \\
\subfloat[Protocol (A) noiseless (MVM)]{\includegraphics[width=0.33\textwidth]{MVM_A-eps-converted-to.pdf}
\label{fig: MVM A}}
\subfloat[Protocol (B) noisy (MVM)]{\includegraphics[width=0.33\textwidth]{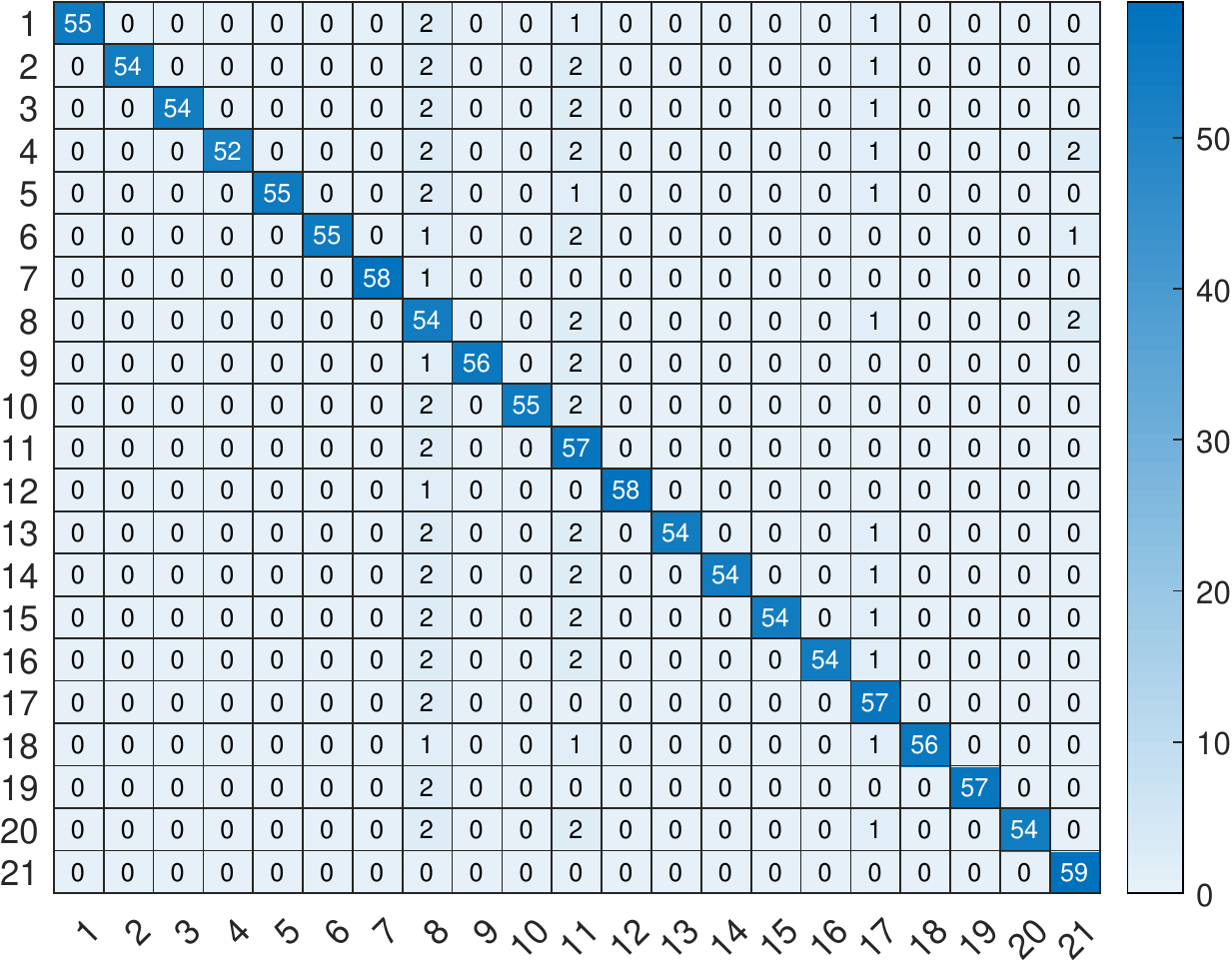}
\label{fig: MVM B}}
\subfloat[Protocol (C) noisy different monitor (MVM)]{\includegraphics[width=0.33\textwidth]{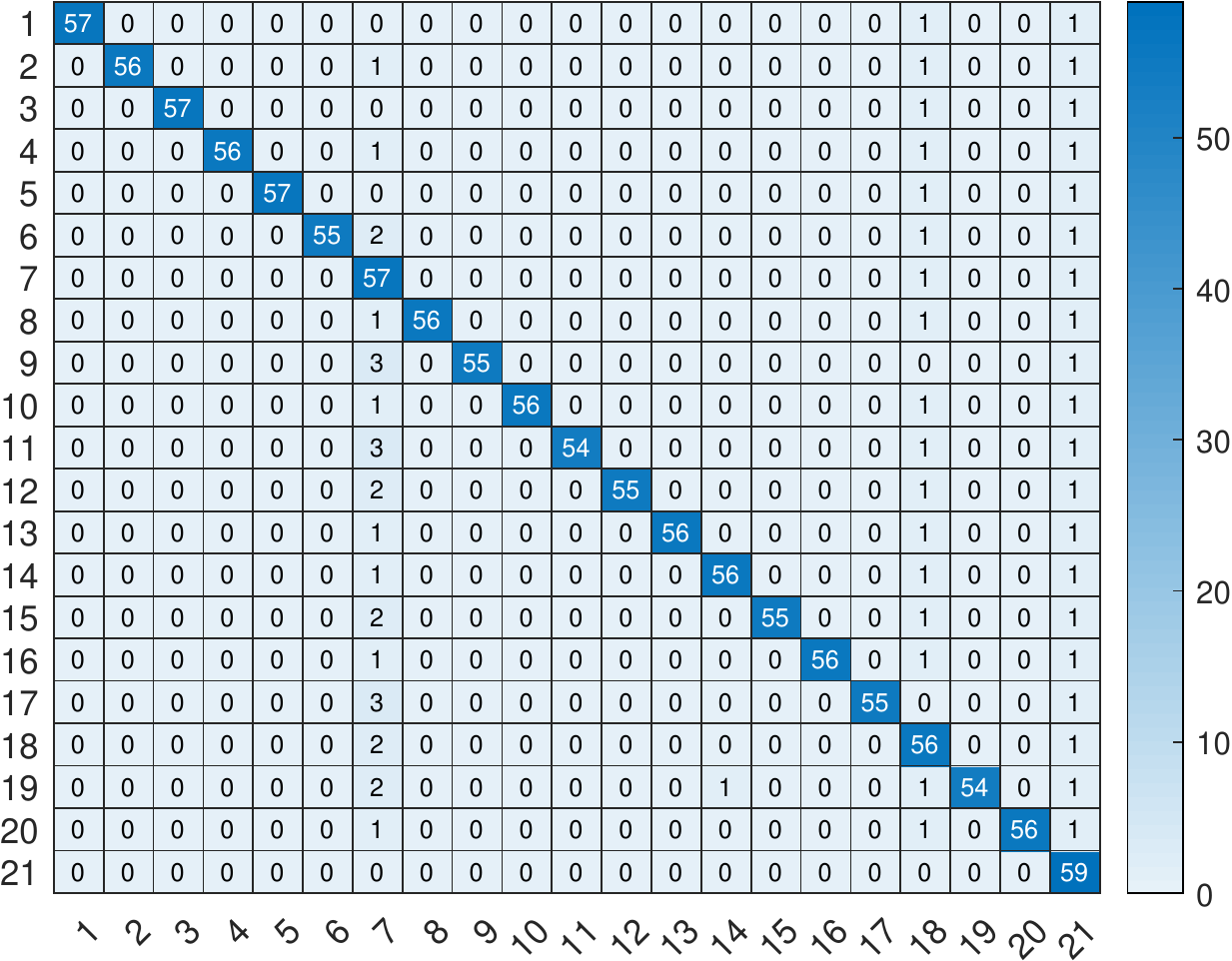}
\label{fig: MVM C}}
\caption{Confusion matrix for three different experimental protocols using DTW and MVM as elastic matching algorithms.}
\label{fig_sim}
\end{figure*}

In a further step the results of the best performing architecture (MVM) were evaluated using confusion matrices for the 21 signals and compared to the baseline system when utilizing DTW. First, when considering protocol A, there is no difference in terms of classification between DTW and MVM as all signals are perfectly classified for each of the 59 noise scenarios (Fig. \ref{fig: DTW A} and Fig. \ref{fig: MVM A}). Second, for protocol B, there is a significant drop in performance for DTW, which is mostly due to misclassification of TV signals with the case of not watching TV (Fig. \ref{fig: DTW B} and Fig. \ref{fig: MVM B}). In detail, there were 35 misclassifications when utilizing DTW, while there have been only 5 misclassifications when utilizing MVM. Third, in protocol C, a similar behaviour is observed as in protocol B with a high number of misclassifications and thus a relatively large performance decrease for DTW and only a small performance decrease for MVM (Fig. \ref{fig: DTW C} and Fig. \ref{fig: MVM C}). In detail, there have been 81 misclassifications with the scenario of not watching TV when utilizing DTW as elastic matching algorithm and 20 misclassifications for MVM respectively. The results of protocol B and C are in line with the work presented in \cite{Latecki.2005} where significantly better performances for elastic matching were reported in noisy scenarios when utilizing MVM. Furthermore, a similar behaviour was reported in our previous study where MVM has also been proven to enhance accuracy for the Energy disaggregation task \cite{Schirmer.2020}.

\section{Conclusion}\label{sec: Conclusion}
In this paper we investigated the potential of identifying the watched TV channel or multimedia content using a smart meter installed outside a house by comparing the household’s aggregated energy consumption signal with a closed set of reference signals acquired from the energy consumption of other monitor devices. The experimental results for a set of 20 possible TV channel options revealed that this is possible and the best identification performance, achieved by the $MVM$ elastic matching algorithm, was equal to 93.8\% in terms of accuracy and 93.3\% in terms of $F_1$ score.

\section*{Acknowledgment}
This work was supported by the UA Doctoral Training Alliance (https://www.unialliance.ac.uk/) for Energy in the United Kingdom. Furthermore, the authors thank Dr Stelios Koutroubinas and Meazon S.A. for providing them with a smart meter device.

% or


\appendices

%\section{}

%\section*{Acknowledgment}


\ifCLASSOPTIONcaptionsoff
  \newpage
\fi


%\IEEEtriggeratref{35}
\bibliographystyle{IEEEtran}
\bibliography{IEEEabrv,Bibliography}\ %IEEEabrv instead of IEEEfull

\end{document}